\documentclass[twocolumn,twoside,slac_two]{revtex4}
\usepackage{graphicx}
\usepackage{fancyhdr}
\newcommand{\etal}{et~al.}

\begin{document}

\title{Shot Analysis of Kepler Blazar W2R~1926$+$42}

%

\author{M. Sasada, S. Mineshige}
\affiliation{Department of Astronomy, Graduate School of Science, Kyoto University, Kitashirakawa-Oiwake-cho, Sakyo-ku, Kyoto 606-8502, Japan; sasada@kusastro.kyoto-u.ac.jp}
\author{S. Yamada}
\affiliation{Department of Physics, Tokyo Metropolitan University, Minami-Osawa 1-1 Hachioji, Tokyo 192-0397, Japan}
\author{H. Negoro}
\affiliation{Department of Physics, Nihon University, 1-8 Kanda-Surugadai, Chiyoda-ku, Tokyo 101-8308, Japan}

\begin{abstract}
Blazars show rapid and violent variabilities, which timescale are often
 less than a day. We studied intraday variations by applying a ``shot
 analysis'' technique to {\it Kepler} monitoring of blazar
 W2R~1926$+$42 in Quarter~14. We obtained a mean profile calculated from
 195 rapid variations. The mean profile shows three components; one is a
 sharp structure distributed within $\pm$0.1~day of the peak, and two
 slow-varying components. This spiky-peak component reflects features of
 rapid variations directly. The profile of peak component shows an
 exponential rise and decay of which timescales are different, 0.0416
 and 0.0588~day respectively. This component is too sharp to represent a
 standard function which is often used to express blazar
 variations. This asymmetric profile at the peak is difficult to be 
 explained by a simple variation of the Doppler factor by changing a
 geometry of the emitting region. This result indicates that intraday
 variations arise from a production of high-energy accelerated particles
 in the jet.

\end{abstract}

\maketitle

\thispagestyle{fancy}


\section{INTRODUCTION}
Blazars have relativistic jets whose axes are closely directed along their
line of sight \citep{Blandford79,Antonucci93}. 
Timescales of brightness variations in blazars are related to sizes of
emitting regions and these speeds in relativistic 
jets. Variations, however, have a variety of timescales from minutes 
to decades. A power spectrum density (PSD) of blazar shows a power-law
distribution, which means that variations of blazars show noise-like
behaviors \citep{Kataoka01}. These various brightness variations in
blazars can be happened by a variety of physical situations in
relativistic jets. The shorter-timescale variations should be reflected
to the physics of inner-emitting regions of a jet. Thus, the study of
short-timescale variations is important to investigate the origin of
variation in blazar jets. 

Blazars show rapid variations having a timescale of less than
1~day. These rapid variations have been reported in wide wavelengths
from radio to gamma-ray bands; in the radio \citep{Quirrenbach92},
optical \citep{Carini90}, X-ray \citep{Kataoka01}, and TeV gamma-ray
bands \citep{Aharonian07}. The {\it Fermi} space telescope has scanned
the entire gamma-ray sky every hours, and detected a lot of large-amplitude
variations as flares \citep{Abdo11,Nalewajko13,Saito13}. Detected flares
in the gamma-ray band often continued for less than 1~day, and had a
variety of shapes, not only a simple rise and decay. We need higher
time-resolution and photon-statistics observations to study the detailed
feature of rapid variations. 

An optical continuous monitoring of blazar W2R~1926$+$42 with a high
time-sampling rate by {\it Kepler} spacecraft \citep{Borucki10} detected
a lot of rapid variations. Its light curve revealed detailed shapes of
numerous variations with large signal-to-noise ratio. We report
general features of rapid variations having a timescale less than 1~day
by stacking these detected variations, and producing a mean profile of rapid
variations, so-called shot analysis. This paper is organized as
follows. Details of {\it Kepler} observation and the way of shot
analysis are described in section~2. Observational results and features
of the mean profile of rapid variations are reported in
section~3. We discuss an origin of rapid variations of the object in
section~4, and section~5 gives several concluding remarks.

\section{OBSERVATION AND ANALYSIS}

\subsection{{\it Kepler} data}
{\it Kepler} monitored over a hundred thousand objects in Cygnus
regions, and obtained continuous light curves with two timing settings,
long (thirty-minute) or short (one-minute) cadences. Blazar
W2R~1926$+$42 is listed in {\it Kepler} target list. It has been
obtained a continuous light curve with the long cadence since
Quarter~11. In Quarter~14, the object had been monitored in the short
cadence mode for 100~days. We produced the calibrated ``SAP\_FLUX''
light curve with one-minute time resolution by the automated 
{\it Kepler} data processing pipeline \citep{Jenkins10}.  

W2R~1926$+$42 is classified as a low-frequency peaked BL Lac object at
$z=0.154$ estimated from two absorption lines
\citep{Edelson12}. Edelson~{\etal} reported that there were numerous
flares with timescales as short as a day in the {\it Kepler} light curve
of Quarter~11 and 12 \cite{Edelson13}. The PSD calculated from the light
curve showed approximately a power-law distribution, but not simple. It
showed a flattening at frequencies below 7$\times$10$^{-5}$~Hz.

\subsection{Shot analysis}
Frequency-domain analyses (e.g. PSD) are not easy to relate with
physical mechanisms directly. On the other hand, time-domain analyses
keeping phase information of variations can be useful for studying
physical mechanisms of variation. We need, however, large photon
statistics to study variations with these time-domain analyses, because
it is difficult to study detailed features of variations by using only
partial data with observational uncertainty. Additionally, observed
variations in blazars usually have a variety of shapes. Thus, it is
difficult to understand general features of variations in blazars by
studing only individual variations.

We apply the light curve of W2R~1926$+$42 obtained by {\it Kepler} to a
shot analysis proposed by Negoro~{\etal} to study the general features
of rapid variations without local features of individual variations,
because a mean profile of rapid variations calculated by the shot
analysis can be cancelled the local features of individual
variations \cite{Negoro94}. We analyzed following procedures to make a
mean profile; First, we select rapid variations as candidates of
shots. Second, we estimate the observational uncertainty in the light
curve. Then, we select the rapid variations with four times larger
amplitudes than the standard deviation of the observational uncertainty
after subtracting the baseline components. We define these variations as
shots.

There are two possibilities for varying the observed brightness,
intrinsic variation of the object and variation by the observational
uncertainty. It is natural that the variation by the observational
uncertainty is dominant rather than the intrinsic variation in shorter
period, especially in the period between two observing points which lie
next to each other. We estimate the standard deviation $\sigma$ of
differences between two neighboring points, and define $\sigma$ as the
observational uncertainty, $\sigma=$17.15 count~s$^{-1}$. 

Rapid variations are often superposed on long-term variations in
light curves of blazars \citep{Sasada08}. We approximate a baseline
component of rapid variation by a second-order local polynomial
fitting to the light curve without the period of the rapid
variation. We subtract the calculated baseline component from
the light curve, and extract the rapid variation. We detect a shot
when the estimated amplitude of extracted rapid variation without the
contribution of the baseline component is larger than our threshold,
$>$4$\sigma$. Additionally, the peak time of the shot is defined at the
time of the maximum flux among the period of rapid variation after
subtracting the baseline component. We calculate a mean profile of
detected shots by stacking with reference to each peak.

\section{RESULTS}

\begin{figure*}[t]
\centering
\includegraphics[width=140mm]{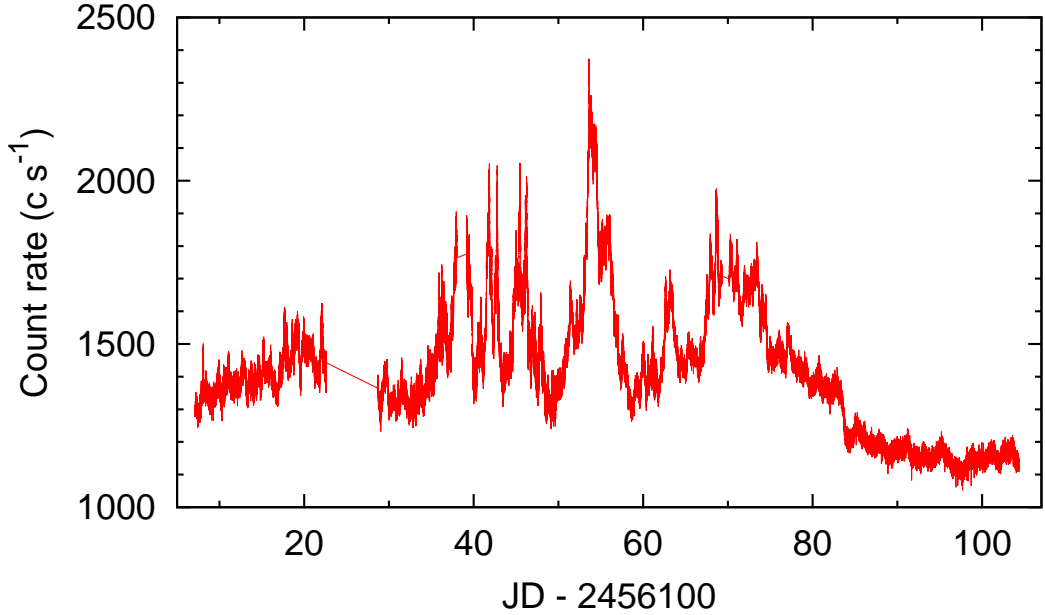}
\caption{Light curves obtained by {\it Kepler} spacecraft in Quarter~14. The
 object monitored for 100~d with one-minute time resolution.}
\label{lc}
\end{figure*}

\begin{figure}[t]
\centering
\includegraphics[width=85mm]{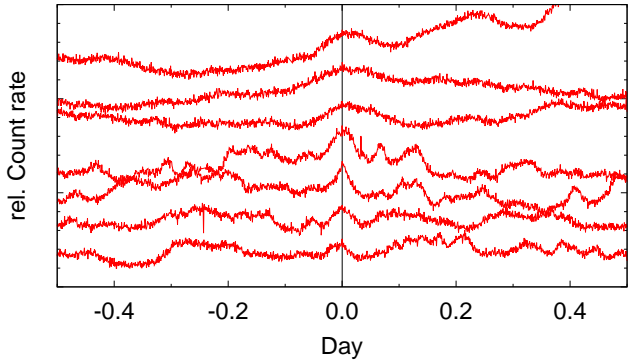}
\caption{Examples of rapid variations. These variations are ranged
 with reference to peak times of individual extrema during the period of
 rapid variations. From bottom to top, the peak times of rapid
 variations are JD~2456134.84, 2456147.23, 2456147.91, 2456151.40,
 2456152.15, and 2456153.04, respectively.}
\label{vari}
\end{figure}

Figure~\ref{lc} shows an optical light curve of the object obtained by
{\it Kepler}. The object showed a violent variability with
various timescales ranging from several tens of minutes to over ten
days, limiting for the observational uncertainty. In the light curve,
there are not only a large-amplitude long-term variation
like from JD~2456150 to 2456160, but also a lot of flare-like
variations with timescale of hours. These rapid variations existed
throughout the entire period of this monitoring. Figure \ref{vari} shows
examples of rapid variations. We range these rapid variations with
reference to the peak times of individual extrema. Figure~\ref{vari}
clearly shows that these variations had a variety of shapes.

\begin{figure}[t]
\centering
\includegraphics[width=85mm]{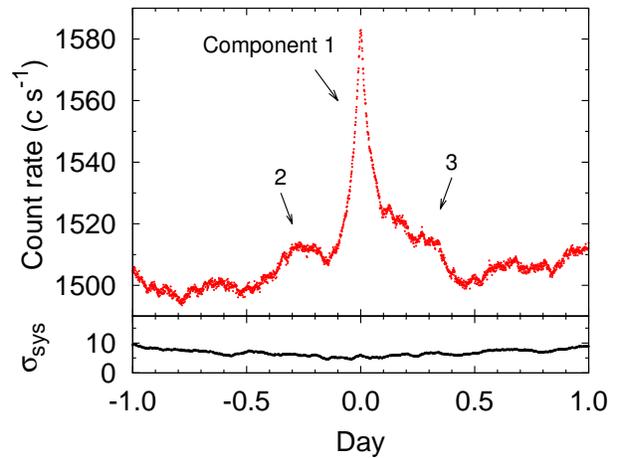}
\caption{A mean profile of detected shots. Upper panel shows the mean
 profile of shots and bottom panel shows the standard deviation
 estimated from a non-parametric bootstrap method. See text for detail.}
\label{shot}
\end{figure}

We detect 195 shots from the obtained light curve, in pursuance of the
definition of shot described in section~2.2. We calculate a mean profile
of these detected shots by appling the shot analysis. Upper panel of
figure~\ref{shot} shows the mean profile of shots without the data at
peak time, because positive fluctuations of the counts at $t=0$ are
summed up systematically \citep{Negoro94}. There are mainly three
components at the mean profile of shots shown in figure~\ref{shot}; a
sharp component distributed in $\pm$0.1~day of the peak time
(component~1), and slow-variable components ranging from $-$0.50 to
$-$0.15~day and from 0.10 to 0.45~day (component~2 and 3),
respectively. An increase and decrease of flux in component~1 are
approximately exponential rise and decay. Additionally, the profile at
the peak is changing from rising to decaying phases for approximately
ten minutes. 

If shot profiles change depending on selected amplitudes of shots, the
calculated mean profile does not reflect to general features of 
shot. We verify whether there is an amplitude dependence to the profile
or not. We separate detected shots in three terms based on these 
amplitudes, 4--6$\sigma$, 6--8$\sigma$, and over 8$\sigma$, and
calculate mean profiles using selected shots. Although calculated
profiles show small differences caused by the limited number of shot
samples, the mean profile of shots has no clear trend associated with
these amplitudes.

We estimate the systematic uncertainty of the obtained mean profile of
shots associated with limited sampling by a non-parametric bootstrap
approach. First, we resample 195 shots with replacement from detected
shots, and calculate a mean profile from the resampled 195 shots. We
produce 10000 pseudo mean profiles of shots following this procedure
with different resamplings. We normalize an average of each mean profile
within $\pm$1~day, and calculate standard deviations in each time
bin. The standard deviations of normalized mean profiles can be regarded
as the systematic uncertainties associated with the sampling of
shots. The bottom panel of figure~\ref{shot} shows calculated standard
deviations. These deviations are ranging from 6 to 10
count~s$^{-1}$. Detected components from 1 to 3 in the mean profile of
shots can be regarded as the real phenomena, not the artificial ones
caused by the systematic uncertainty of the sampling of shots.

\begin{figure}[t]
\centering
\includegraphics[width=85mm]{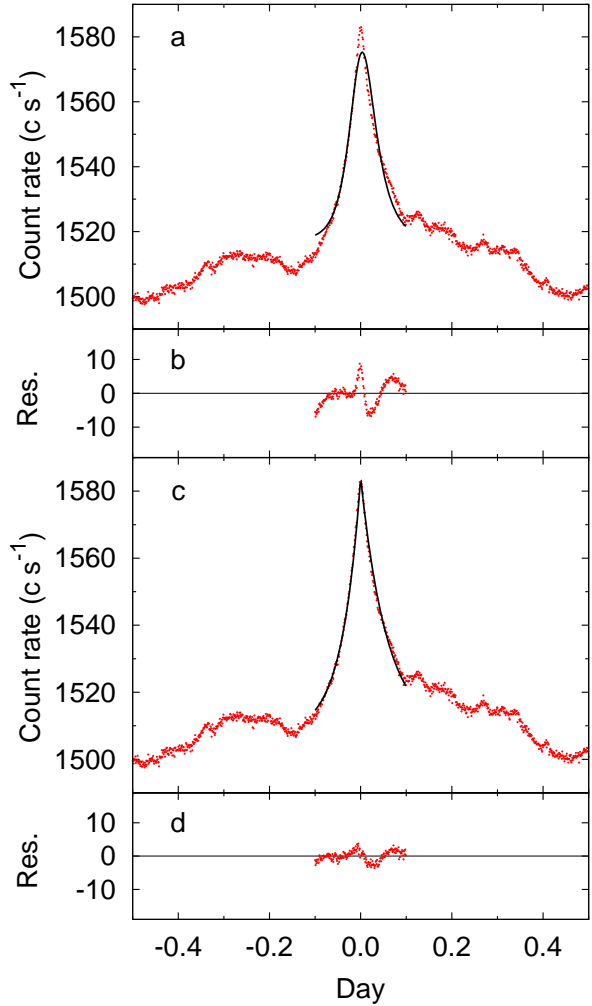}
\caption{Best-fitted functions superposed on mean profiles of
 shots. Panels ``a'' and ``c'' show a mean profile of shots and the
 best-fitted functions~(\ref{eq:con}) and (\ref{eq:exp}). Panels ``b'' and
 ``d'' show residuals between the mean profiles and the estimated
 best-fitted functions.}
\label{fit}
\end{figure}

Component~1 reflects general features of shots directly, because this
component is distributed around the peak time. First, we apply this
exponential shape of component~1 to a function proposed by Abdo~{\etal}; 
\begin{eqnarray}
F(t) = F_{0}\;[e^{-t/T'_{r}}+e^{t/T'_{d}}]^{-1} + F_{c}, \label{eq:con}
\end{eqnarray}
where $T'_{r}$ and $T'_{d}$ are variation timescales of rise and
decay phases, $F_{c}$ represents a constant level underlying
the component~1, and $F_{0}$ measures the amplitude of the shot
\cite{Abdo10}. We evaluate its goodness of fit by ${\chi}^{2}$ test,
${\chi}^{2}=\sum\left\{F_{\rm data}(t_{i})-F_{\rm model}(t_{i})\right\}^2$.
The ${\chi}^2$ of the best fitted function is 2278 within $\pm$0.1~day
of the mean profile to except contaminations of other components. On
the other hand, we apply an another function;  
\begin{eqnarray}
F(t) = \left\{ 
\begin{array}{ll}
F_{0}\; e^{-t/T_{r}} + F_{c},\; & t < 0 \\
F_{0}\; e^{t/T_{d}} + F_{c}, & t > 0, \label{eq:exp}
\end{array}
\right.
\end{eqnarray}
where $T_{r}$ and $T_{d}$ are e-folding times of rise and decay, and
$F_{c}$ and $F_{0}$ are the same in the case of
function~(\ref{eq:con}). We also calculated the $\chi^2$ of this function,
$\chi^2$=469. Figure~\ref{fit} shows the applied functions with the best
fitted parameters superposed on the mean profile and its residuals.
Although the function~(\ref{eq:con}) shows obvious residuals during the
peak time in panels ''a'' and ``b'' of figure~\ref{fit}, the
residuals in the case of function~(\ref{eq:exp}) are suppressed shown in
panels ``c'' and ``d''. This indicates that the mean profile is more
spiky than the expected profile from function~(\ref{eq:con}). Therefore,
the goodness of fit of the function~(\ref{eq:exp}) is more plausible
than that of function~(\ref{eq:con}) to represent the component~1 of the
mean profile. 

We estimated the best-fitted parameters with a chi-squared test. The mean
profile, however, has a systematic uncertainty caused by the sampling of
shots as mentioned above. We applied the non-parametric bootstrap
approach to calculate the confidence level with the same way to estimate
the errors of calculated parameters. First, we calculated 10000 pseudo
mean profiles estimated from resamples with replacement from detected
shots. We calculated the best-fitted parameters against individual
pseudo profiles, and estimated the confidence levels in these
parameters. In table~1, we show the best-fitted parameters of
function~(\ref{eq:exp}) to the mean profile and the ranges of 95~\%
confidence levels. We applied the Wilcoxon rank-sum test which was a
non-parametric significance test (also referred to as the Mann-Whitney
U-test) to the distributions of $T_{r}$ and $T_{d}$ calculated by the
bootstrap approach \citep{Wilcoxon45,Mann47}. We confirmed the
difference between the $T_{r}$ and $T_{d}$, because the p-value was less
than $10^{-15}$. Therefore, the component~1 in the mean profile of shots
has an asymmetric profile.

\begin{table}[t]
\begin{center}
\caption{Parameters of best-fitted function~(\ref{eq:exp}) to
 component~1 of the mean profile of shots \label{table:fit}}
\begin{tabular}{|l|c|c|}
\hline \textbf{} & \textbf{Best value} & \textbf{95\% confidence level} 
\\
\hline $T_{r}$ (day) & 0.0416 & [0.0320, 0.0543] \\
\hline $T_{d}$ (day) & 0.0588 & [0.0399, 0.0919] \\
\hline $F_{0}$ (count~s$^{-1}$) & 76 & [65, 88] \\
\hline $F_{c}$ (count~s$^{-1}$) & 1508 & [1484, 1537] \\
\hline
\end{tabular}
\label{table}
\end{center}
\end{table}

\section{DISCUSSION}
We obtained the optical continuous light curve of blazar W2R~19426$+$42
with one-minute time resolution by {\it Kepler} spacecraft. The object
showed violent variability and a lot of rapid variations with timescales
less than a day. We detected 195 rapid variations as shots of which
amplitude were larger than 4$\sigma$ after subtracting these baseline
components, and applied to the shot analysis. The mean profile produced from
detected shots shows three components, one fast-spiky (component~1) and
two slow-varying components. Component~1 shows an asymmetric profile,
faster-rise and slower-decay features with the spiky but
smooth-connected peak.

It is poorly understood whether rapid variations are intrinsic phenomena
or apparent one caused by a geometrical changing in the jet. There are
several models that flux variations are explained as apparent brightness
variations, for example varying the Doppler factor for changing the
viewing angle \citep{Villata99} or gravitational lensing effect
\citep{Chang79}. These models, however, expect that the averaged
variation profile is almost symmetric in a simple situation, because the
Doppler factor should be changed symmetrically in the averaged variation.
In other words, the rise and decay timescales should be equal. Estimated
rise and decay timescales of rapid variations, however, are different
described in section~3. Thus, these models can be ruled out in the case
of rapid variations. Thus, rapid variations may come from authentic
phenomena. It is plausible that there is a particle acceleration during
the rapid variation, and then, higher-energy particles increase in the
emitting region of the variation.

The synchrotron cooling timescale ${\tau}_{\rm syn}$ is represented as,
${\tau}_{\rm syn}{\sim}3.2{\times}10^{4}B^{-3/2}E^{-1/2}{\delta}^{-1/2}$~sec, 
where $B$ is a strength of magnetic field, $E$ is an observed energy
band, and ${\delta}$ is the Doppler factor
\citep{Tashiro95,Sasada10}. If the dissipation of high-energy particles
in rapid variations is caused by the synchrotron cooling, 
$\tau_{\rm syn}$ in the rest frame can be represented using the decaying
timescale $T_{d}$, ${\tau}_{\rm syn}={\delta}\;T_{d}/(1+z)$. We estimate
$\delta$ of 5.8 from observed $T_{d}$ of the mean profile of shots, where
$E$ is 2.25~eV and assuming $B$ as 0.5~G which is typical value among
the gamma-ray detected BL~Lac objects \citep{Ghisellini10}. The mean
profile reflects common features of rapid variations. Thus, the
estimated $\delta$ should be a typical value of inner regions where
rapid variations happen.

\section{CONCLUSION}
The optical continuous light curve with one-minute time sampling obtained
by {\it Kepler} revealed that the mean profile of rapid variations
almost showed exponential rise and decay. Rise and decay timescales of
shot profile, however, are different, and the profile shows asymmetric
profile. A particle acceleration process can produce this asymmetric
variations. There are several scenarios which can explain the
particle-acceleration mechanism causing the rapid variations;
shock-in-jet scenario \citep{Marscher85}, magnetic reconnection scenario
\citep{Giannios09}. The shot analysis is also feasible to study the
spectral feature of variations, because of large signal-to-noise
ratio. Unfortunately, {\it Kepler} performed only one-band
monitoring. Spectral and further observational studies are needed to
completely understand the mechanism of rapid variations.

\bigskip 
\begin{acknowledgments}
This work was supported by a Grant-in-Aid for JSPS Fellows.

\end{acknowledgments}

\bigskip 

\end{document}